\documentclass[journal]{IEEEtran}
\usepackage{graphicx}
\usepackage{xspace}

\newcommand{\nanet}{NaNet\xspace}
\newcommand{\nanetone}{\mbox{NaNet-1}\xspace}
\newcommand{\nanetten}{\mbox{NaNet-10}\xspace}

\newcommand{\gbe}{GbE\xspace}

\newcommand{\nvidia}{NVIDIA\xspace}
\newcommand{\pcie}{PCIe\xspace}

\newcommand{\ie}{\textit{i.e.}\xspace}
\newcommand{\eg}{\textit{e.g.}\xspace}

\newcommand{\cerenkov}{\v{C}erenkov\xspace}

\newcommand{\de}{DE5-net\xspace}

\newcommand{\gpudirect} {GPUDirect\xspace}

\newcommand{\GPULZEROTP} {GPU\_L0TP\xspace}

\graphicspath{{fig/}}
\DeclareGraphicsExtensions{.pdf,.jpeg,.png}

\begin{document}
\title{GPU-based Real-time Triggering in the NA62 Experiment}

\author{
  R.~Ammendola\IEEEauthorrefmark{2},
  A.~Biagioni\IEEEauthorrefmark{1},
  P.~Cretaro\IEEEauthorrefmark{1},
  S.~Di Lorenzo\IEEEauthorrefmark{3},
  R.~Fantechi\IEEEauthorrefmark{3},
  M.~Fiorini\IEEEauthorrefmark{7},
  O.~Frezza\IEEEauthorrefmark{1},
  G.~Lamanna\IEEEauthorrefmark{3}\IEEEauthorrefmark{4}, 
  F.~Lo~Cicero\IEEEauthorrefmark{1},
  A.~Lonardo\IEEEauthorrefmark{1}
  M.~Martinelli\IEEEauthorrefmark{1},
  I.~Neri\IEEEauthorrefmark{7},
  P.~S.~Paolucci\IEEEauthorrefmark{1}, 
  E.~Pastorelli\IEEEauthorrefmark{1}, 
  R.~Piandani\IEEEauthorrefmark{3},
  L.~Pontisso\IEEEauthorrefmark{3},
  D.~Rossetti\IEEEauthorrefmark{6},
  F.~Simula\IEEEauthorrefmark{1}, 
  M.~Sozzi\IEEEauthorrefmark{3}\IEEEauthorrefmark{4} ,
  and P.~Vicini\IEEEauthorrefmark{1}\\

%% \IEEEmembership{$^a$INFN Sezione di Roma,~Italy}\\
%% \IEEEmembership{$^b$INFN Sezione di Tor Vergata,~Italy}\\
%% \IEEEmembership{$^c$CERN,~Switzerland}\\
%% \IEEEmembership{$^d$NVIDIA Corporation,~U.S.A.}% <-this % stops a space

  \thanks{\IEEEauthorrefmark{1}INFN Sezione di Roma,~Italy.}%
  \thanks{\IEEEauthorrefmark{2}INFN Sezione di Tor Vergata,~Italy.}%
  \thanks{\IEEEauthorrefmark{3}INFN Sezione di Pisa,~Italy.}%
  \thanks{\IEEEauthorrefmark{4}CERN,~Switzerland.}%
  \thanks{\IEEEauthorrefmark{5}Universit\`a di Roma Sapienza, Dipartimento di Fisica,~Italy.}%
  \thanks{\IEEEauthorrefmark{6}NVIDIA Corporation,~U.S.A.}% <-this % stops a space
  %\thanks{M.~Martinelli is the corresponding author (\mbox{michele.martinelli@roma1.infn.it}.)}% <---nel PREPRINT credo non vada messo
  %\thanks{This work was supported in part by the EU Framework Programme 7
  %  EURETILE project, grant number 247846; R. Ammendola was supported by
  %  MIUR (Italy) through the INFN SUMA project; G.~Lamanna and M.~Sozzi thank
  %  the GAP project, partially supported by MIUR under grant RBFR12JF2Z
  %  ``Futuro in ricerca 2012''.}% <-this % stops a space
  %\thanks{Full names of authors are preferred in the author field, but are not required. Put a space between authors' initials. Do not use all uppercase for authors' surnames.}%
  \thanks{\IEEEauthorrefmark{7}INFN Sezione di Ferrara,~Italy.}%
  %\thanks{Manuscript received }%
}

\maketitle
\thispagestyle{empty}

\begin{abstract}

Over the last few years the GPGPU (General-Purpose computing on
Graphics Processing Units) paradigm represented a remarkable
development in the world of computing.
Computing for High-Energy Physics is no exception: several 
works have demonstrated the effectiveness of the integration 
of GPU-based systems in high level trigger of different experiments.
On the other hand the use of GPUs in the low level trigger systems,
characterized by stringent real-time constraints, such as tight time
budget and high throughput, poses several challenges.
In this paper we focus on the low level trigger in the CERN NA62
experiment, investigating the use of real-time computing on GPUs in 
this synchronous system. 
Our approach aimed at harvesting the GPU computing power to build
in real-time refined physics-related trigger primitives for the RICH detector, 
as the the knowledge of Cerenkov rings parameters allows to build stringent 
conditions for data selection at trigger level.
Latencies of all components of the trigger chain have been analyzed, 
pointing out that networking is the most critical one. 
To keep the latency of data transfer task under control, we devised NaNet, 
an FPGA-based PCIe Network Interface Card (NIC) with GPUDirect capabilities. 
For the processing task, we developed specific multiple ring trigger algorithms to
leverage the parallel architecture of GPUs and increase the processing
throughput to keep up with the high event rate.
Results obtained during the first months of 2016 NA62 run are
presented and discussed.

\end{abstract}

%\begin{IEEEkeywords}
%IEEEtran, journal, \LaTeX, paper, template.
%\end{IEEEkeywords}

%% \section{Introduction}
%% %Alessandro, Piero
%% Il problema in termini generali: uso di acceleratori GPU in ambito HEP
%% per sistemi \realtime.  -high latency e jiitter per commodity network
%% -...

\section{Introduction}
\label{sec:intro}

In High Energy Physics experiments the realtime selection of the most interesting events is of paramount importance because of the collision rates which do not give the possibility to save all the data for offline analysis. For this purpose, different trigger levels are usually used to carefully choose the most meaningful events.
The low level ones require low and (almost) deterministic latency and their standard implementation is on dedicated hardware (ASICs or FPGAs).
Our approach aims at exploiting the Graphic Processing Units (GPUs)
computing power, in order to build refined \mbox{physics-related}
trigger primitives, such as energy or direction of the final state
particles in the detectors, and therefore leading to a net improvement 
of trigger conditions and data handling.
GPUs architectures are massively parallel, being designed to optimize
computing throughput but with no particular attention to their usage
in real-time contexts, such as the online low level triggers. While
execution times are rather stable on these devices, also I/O tasks
have to guarantee real-time features along the data stream path, from
detectors to GPU memories.
The NaNet project arises with the goal of designing a low-latency and high-throughput data transport mechanism for systems based on CPU/GPUs.
The GPU-based L0 trigger using the NaNet board is currently integrated in the experimental setup of the RICH \cerenkov detector of the NA62 experiment in order to reconstruct the ring-shaped hit patterns.
We report and discuss results obtained with this system along with the algorithms that will be implemented.

\section{\nanet architecture}
\label{sec:nanet}

The design of a \mbox{low-latency}, \mbox{high-throughput} data
transport mechanism for \mbox{real-time} systems is mandatory in order
to bridge the \mbox{front-end} electronics and the software
trigger computing nodes~\cite{NanetTweppShort:2014} of High Energy Physics Experiments.
\nanet, being an FPGA\-based NIC, natively supports a variety of link
technologies allowing for a straightforward integration in different
experimental setups.
Its key characteristic is the management of custom and standard
network protocols in hardware, in order to avoid OS jitter effects and
guarantee a deterministic behaviour of communication latency while
achieving maximum capability of the adopted channel.
Furthermore, \nanet integrates a processing stage which is able to
reorganize data coming from detectors on the fly, in order to improve
the efficiency of applications running on computing nodes.
On a \mbox{per-experiment} basis, different solutions can be
implemented: data decompression, reformatting, merging of event
fragments.

Finally, data transfers to or from application memory are directly
managed avoiding bounce buffers.
\nanet accomplishes this \mbox{zero-copy} networking by means of a
hardware implemented memory copy engine that follows the RDMA paradigm
for both CPU and GPU --- this latter supporting the \gpudirect V2/RDMA
by \nvidia to minimize the I/O latency in communicating with GPU
accelerators.

On the host side, a dedicated Linux kernel driver offers its services
to an application level library, which provides the user with a series
of functions to: open/close the \nanet device; register and
\mbox{de-register} circular lists of persistent data receiving buffers
(CLOPs) in GPU and/or host memory; manage software events generated
when a receiving CLOP buffer is full (or when a configurable timeout
is reached) and received data are ready to be processed.

\nanetone was developed in order to verify the feasibility of the
project; it is a \pcie Gen2 x8 network interface card featuring
\gpudirect RDMA over \gbe.

\nanetten is a \pcie Gen2 x8 network adapter implemented on the
Terasic \de board equipped with an Altera Stratix V FPGA featuring
four SFP+ cages~\cite{NanetTwepp:2015}.

Both implementations use UDP as transport protocol.

\begin{figure}[!hbt]
\centering
  \includegraphics[trim=10mm 0mm 10mm 30mm,clip,width=0.49\textwidth]{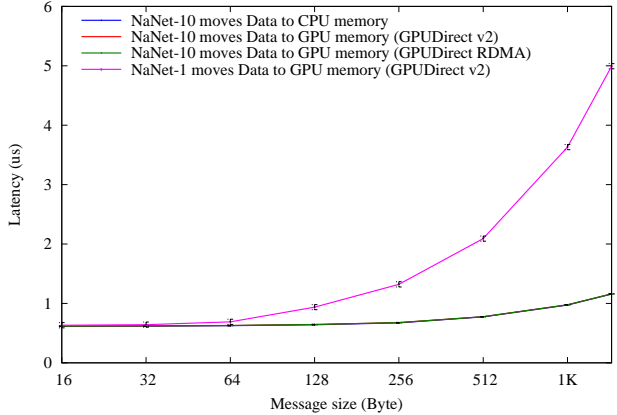}
  \caption{\nanetten vs. \nanetone hardware latency. \nanetten curves are completely overlapping at this scale.}
  \label{fig:latency}
\end{figure}

\begin{figure}[!hbt]
\centering
  \includegraphics[trim=10mm 0mm 10mm 30mm,clip,width=0.49\textwidth]{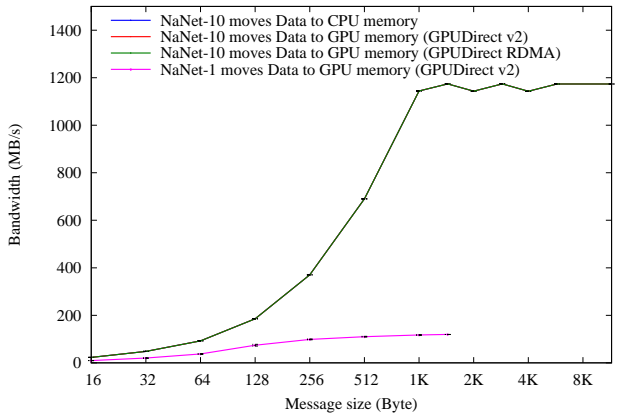}
  \caption{\nanetten vs. \nanetone bandwidth. \nanetten curves are completely overlapping at this scale.}
  \label{fig:bandwidth}
\end{figure}

In Fig.~\ref{fig:latency}, \nanetten and \nanetone latencies are
compared within UDP datagram size range; \nanetten guarantees
\mbox{sub-$\mu$s} hardware latency for buffers up to $\sim$1kByte in
GPU/CPU and it reaches its 10~Gbps bandwidth peak capability already
at $\sim$1kByte size (Fig.~\ref{fig:bandwidth}).

\section{NA62 experiment}
\label{sec:na62experiment}

The NA62 experiment at CERN~\cite{Lamanna:2011zz} aims at measuring
the Branching Ratio of the \mbox{ultra-rare} decay of the charged 
Kaon into a pion and a \mbox{neutrino-antineutrino} pair.
The NA62 goal is to collect $\sim100$ events with a signal to 
background ratio 10:1, using a novel technique with a \mbox{high-energy} 
(75~GeV) unseparated hadron beam decaying in flight.
In order to manage the \mbox{high-rate} data stream due to a
$\sim$ 10~MHz rate of particle decays illuminating the detectors,
a set of trigger levels will have to reduce this rate by three
orders of magnitude.
The entire trigger chain works on the main digitized data
stream~\cite{Avanzini:2010zz}.

The Low-level trigger, implemented in hardware by means of FPGAs
on the readout boards, reduces the data stream by a factor 10 to meet
the maximum design rate for event readout of 1~MHz.
The upper trigger levels (L1 and L2) are 
\mbox{software-implemented} on a commodity PC farm for further
reconstruction and event building. 

In the standard implementation, the FPGAs on the readout boards
compute simple trigger primitives on the fly, such as hit
multiplicities and rough hit patterns, which are then
\mbox{time-stamped} and sent to a central processor for matching and
trigger decision.
Thus the maximum latency allowed for the synchronous L0 trigger is related
to the maximum data storage time available on the data acquisition
boards.
For NA62 this value is up to 1~ms, in principle allowing use of more
compute demanding implementations at this level, \ie the GPUs.

\section{Implementation of the GPU-based Low-level trigger}
\label{sec:GPUbasedtrigger}

\begin{figure}[!hbt]
\centering 
%  \includegraphics[trim=0mm 40mm 0mm 40mm,clip,width=\textwidth]{fL0gpu.pdf}
%\textwidth
  \includegraphics[trim=0mm 40mm 0mm 40mm,clip,width=0.55\textwidth]{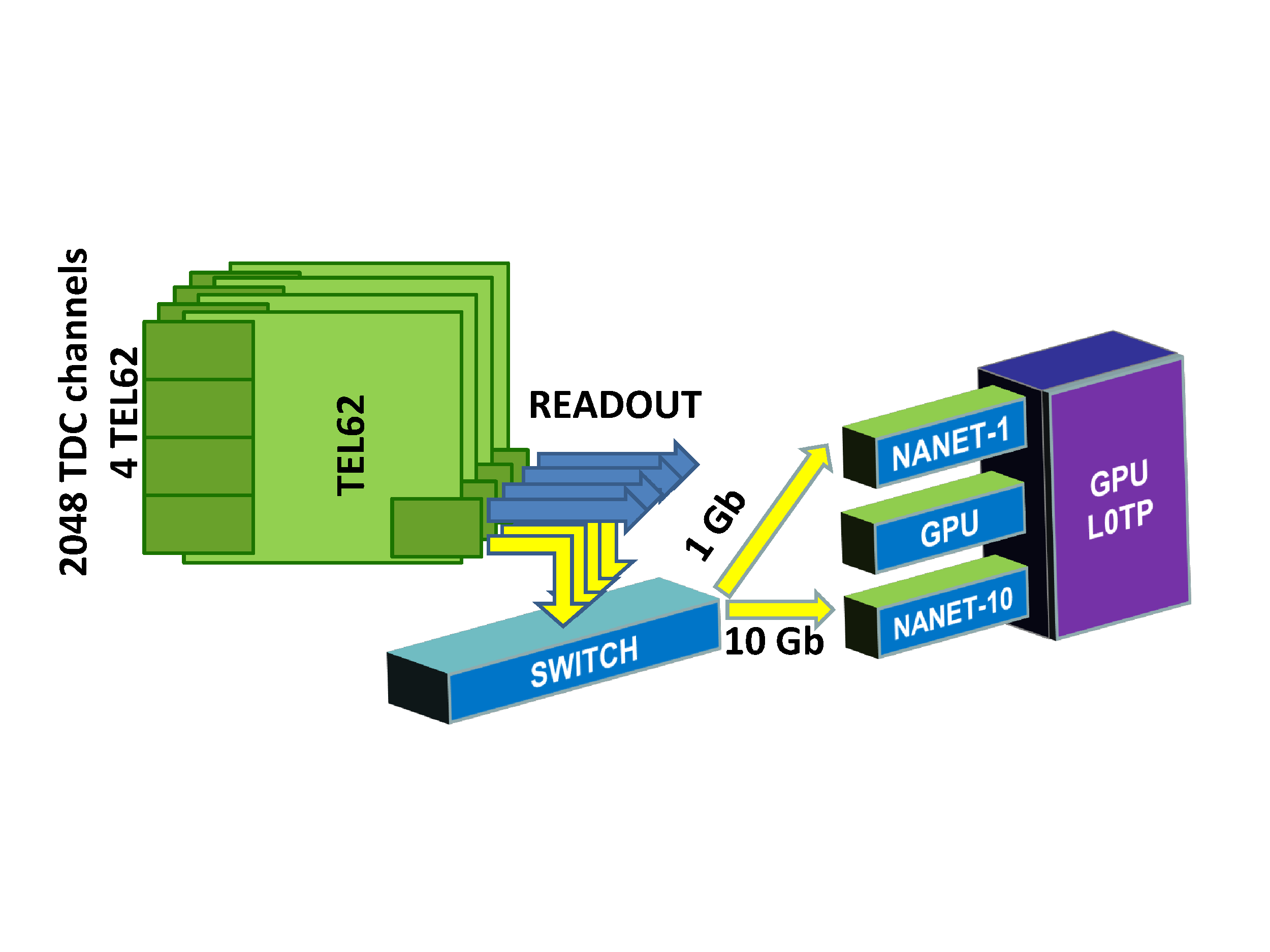}
  \caption{Pictorial view of \mbox{GPU-based} Trigger.}
  \label{fig:gputrigger}
\end{figure}

As a first example of GPU application in the NA62 trigger system we
studied the possibility to reconstruct rings in the RICH.
This detector identifies pions and muons with momentum in the range between
15~GeV/c and 35~GeV/c.
\cerenkov light is reflected by a composite mirror with a focal length
of 17~m focused onto two separated spots equipped with $\sim
1000$~photomultipliers (PM) each.
Data communication between the readout boards (TEL62) and
the L0 trigger processor happens over multiple \gbe links using UDP
streams.
The final system consists of 4~\gbe links to move primitives data from
the readout boards to the \GPULZEROTP (see Fig.~\ref{fig:gputrigger}).
The overall time budget for the low level trigger comprising both communication and computation tasks is of 1~ms, so a deterministic response latency from \GPULZEROTP is a strict requirement.
Refined primitives coming from the GPU-based calculation will be then sent to the central L0 processor, where the trigger decision is made taking in account informations from other detectors.

\subsection{Multiple ring events recontruction on GPU}
\label{sec:computing}

Taking the parameters of \cerenkov rings into account could be very useful in order to build stringent conditions for data selection at trigger level.
This implies that circles have to be reconstructed using the coordinates of activated PMs.

We take in consideration two multi-rings pattern recognition algorithms based only on geometrical considerations (no other information is available at this level) and particularly suitable for exploiting the intrinsic parallel architecture of GPUs.

\subsubsection{Histogram algorithm}
\label{sec:histogram}

%The first is a \mbox{histogram-based} algorithm in which the XY plane is divided into a grid and a histogram is created with distances from gridpoints and hits of the physics event. 
The procedure involves dividing the XY plane into a grid and creating a histogram whose bins contain distances from the gridpoints and the hits of the physics event.
Distance bins whose contents exceed a threshold value let identify the rings.
In order to limit the use of resources, it is possible to proceed in two steps, starting the histogram procedure with a 8x8 grid and calculating now distances from such squares.
Afterwards, to refine their positions, the calculation is repeated with a grid 2x2 only for the candidate centers selected according to the threshold in the previous step.

\subsubsection{Almagest algorithm}
\label{sec:almagest}

%The second is based on Ptolemy's Theorem which states that when four vertices of a quadrilateral (ABCD) lie on a common circle, it is possible to relate four sides and two diagonals: $|AC|\times|BD|=|AB|\times|CD|+|BC|\times|AD|$. This formula can be implemented in a parallel way allowing for a fast multi-ring selection.

\begin{figure}[!hbt]
\centering
  \includegraphics[trim=0mm 0mm 0mm 0mm,clip,width=0.45\textwidth]{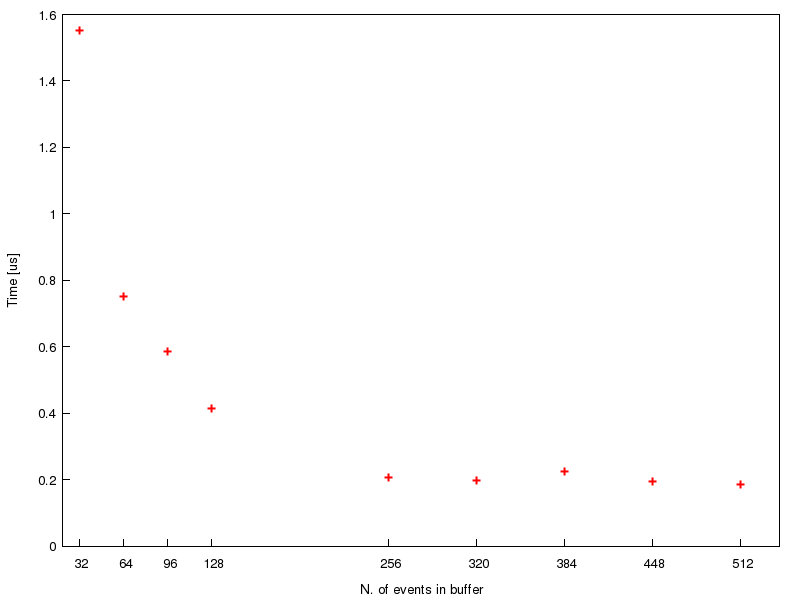}
  \caption{Almagest algorithm performances, time for single event. Test performed on K20c \nvidia GPU.}
  \label{fig:almagesttest}
\end{figure}

The Ptolemy's Theorem states that when four vertices of a quadrilateral (ABCD) lie on a common circle, it is possible to relate four sides and
two diagonals: $|AC|\times|BD|=|AB|\times|CD|+|BC|\times|AD|$. By using this formula it is possible to implement a pattern recognition algorithm for multi-rings which exposes different level of parallelism, resulting well-suited for GPUs achitecture and fast in its execution.
This is crucial either to directly reconstruct the rings or to choose different algorithms according to the number of circles.
The large number of possible combinations of four vertices, given a maximum of 64 points for physics event, can be a limitation to this approach.
To greatly reduce the number of tests, one possibility is to choose few triplets --- \ie a set of three hits assumed to belong to the same ring --- trying to maximize the probability that all their points belong to the same ring and iterating through all the remaining hits to search for the ones satisfying the aforementioned formula~\cite{Lamanna:almagest}.
The parallel implementation of this algorithm yields many triplets and events being processed at the same time. Some results are shown in Fig.~\ref{fig:almagesttest}, where the computing time is further sped up by greatly reducing accesses to GPU shared memory, mainly using threads private registers through CUDA intra-warp veto and shuffle instructions, so that multi-rings events are processed with a latency of 0.5~us per event.
Once the number of rings and points belonging to them have been found, it is possible to apply \eg Crawford's method~\cite{Crawford:1983td} to obtain centre coordinates and radii with better spatial resolution.

\section{Results}
\label{sec:results}

In 2015 the \mbox{GPU-based} trigger at CERN comprises 2~TEL62
readout boards connected to a HP2920 switch and a \nanetone board with
a TTC HSMC daughtercard plugged into a SuperMicro server consisting of
a \mbox{X9DRG-QF} dual socket motherboard --- Intel C602 Patsburg
chipset --- populated with Intel Xeon \mbox{E5-2620} @2.00~GHz CPUs
(\ie Ivy Bridge \mbox{micro-architecture}), 32~GB of DDR3 memory and
a \mbox{Kepler-class} \nvidia K20c GPU.

Such a system allows for testing of the whole chain: the data events
move towards the \mbox{GPU-based} trigger through \nanetone by means
of the \gpudirect RDMA interface.
Data arriving within a configurable time frame are gathered and then organized in a Circular List Of Persistent
buffers (CLOP) in the GPU memory.
Buffer number and size are tunable in order to optimize computing and
communication.
This time frame must obviously be shorter or equal on average to how long the GPU takes for \mbox{multi-ring} reconstruction, to be sure
that buffers are not overwritten by incoming events before they are consumed by the GPU.
Events coming from different TEL62 need to be merged in the GPU memory
before the launch of the ring reconstruction kernel.
Each event is timestamped and the ones coming from different readout
boards that are in the same \mbox{time-window} are fused in a single
event describing the status of PMs in the RICH detector.

\begin{figure}
\centering
  \includegraphics[trim=0mm 0mm 0mm 0mm,clip,width=0.49\textwidth]{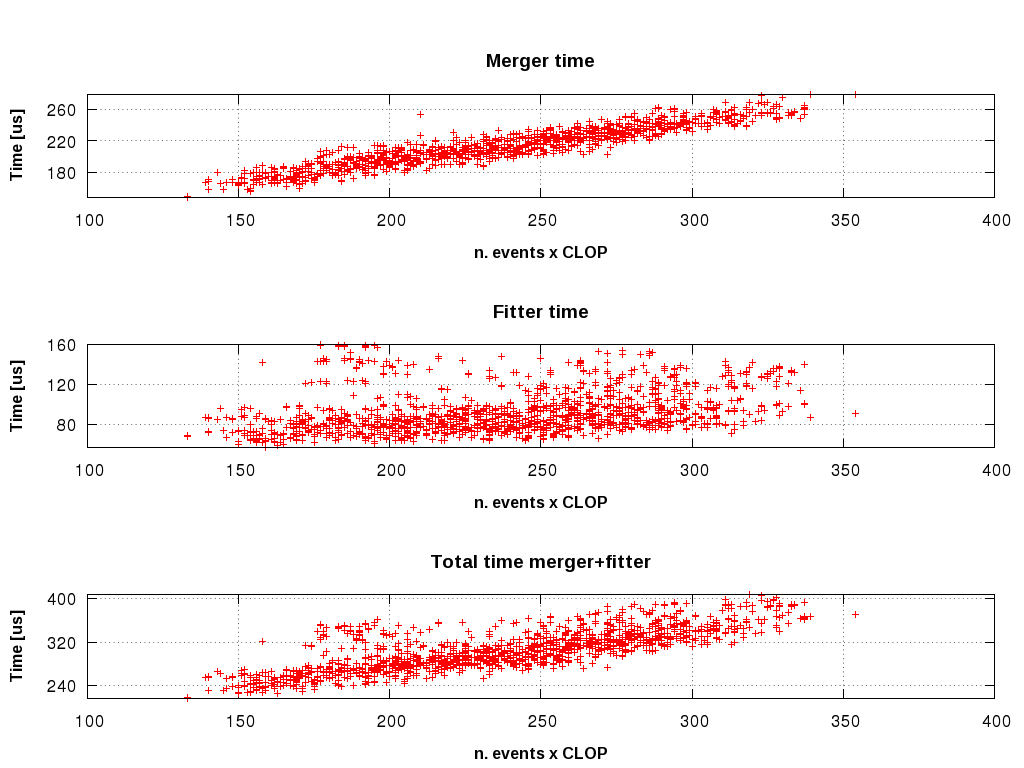}
  \caption{\mbox{Multi-ring} reconstruction of events performed on K20c \nvidia GPU.}
  \label{fig:trigresults}
\end{figure}

Results are reported in Fig.~\ref{fig:trigresults}. The CLOP size measured as number of received events is on the \mbox{X-axis} and the latencies of different stages are on the \mbox{Y-axis}. The computing kernel implemented the histogram fitter with a single step (\ie using an 8x8 grid only). Events coming from 2 readout boards, for a gathering time of 400~$\mu$s, and parameters like events rate (collected with a beam intensity of $4\times10^{11}$~protons per spill), a CLOP's size of 8KB, time frame was chosen so that we could test the online behaviour of the trigger chain.

Because the merge operation doesn't expose much parallelism, requiring instead synchronization and serialization, this is an \mbox{ill-suited} problem to the GPU architecture. In operative conditions, the merging time only would exceed the time frame. The high latency of the merger task when performed on a GPU strongly suggests to offload such duties to a hardware implementation.

\section{Conclusions and future work}

\section*{Acknowledgment}
S.~Di~Lorenzo, R.~Fantechi, M.~Fiorini, I.~Neri, R.~Piandani, L.~Pontisso, M.~Sozzi thank the GAP project, partially supported by MIUR under grant RBFR12JF2Z ``Futuro in ricerca 2012''.

%% The preferred spelling of the word ``acknowledgment'' in American English is without an ``e'' after the ``g.'' Use the singular heading even if you have many acknowledgments. Avoid the expression, ``One of us (S.B.A.) thanks ...'' Instead, write ``S.B.A. thanks ...'' Put sponsor acknowledgments in the unnumbered footnote on the first page.


\bibliographystyle{IEEEtran}
\bibliography{IEEEabrv,../../../../ape_bib/bibliography.bib}
%% \begin{thebibliography}{2}

%% \bibitem{IEEEhowto:kopka}
%% H.~Kopka and P.~W. Daly, \emph{A Guide to \LaTeX}, 3rd~ed.\hskip 1em plus
%%   0.5em minus 0.4em\relax Harlow, England: Addison-Wesley, 1999.

%% \bibitem{IEEEPDFRequirement401}
%% IEEE Content Engineering, \emph{IEEE PDF Specification Version 4.10}. Available: http://www.ieee.org/documents/31296\_IEEE\_PDF\_Spec.zip.

%% \end{thebibliography}

\end{document}